\documentclass[twocolumn]{revtex4-1} \usepackage{graphicx}
\usepackage{dcolumn}
\usepackage{bm}
\usepackage{amssymb}
\usepackage{latexsym}
\usepackage{tensor}
\begin{document}
\title{ Correspondences of matter fluctuations in semiclassical and classical
gravity for cosmological spacetime } 
\author{ Seema Satin }
\affiliation{ Indian Institute for Science Education and Research, Mohali,
 Punjab}
\email{satin@iisermohali.ac.in}
\newcommand{\be}{\begin{equation}}
\newcommand{\ee}{\end{equation}}
\newcommand{\bea}{\begin{eqnarray}}
\newcommand{\eea}{\end{eqnarray}}
\newcommand{\G}{\hat{G}}
\begin{abstract}
A correspondence between  fluctuations of conformally invariant quantum fields
 and that of classical fields finally reducing to perfect fluid matter 
content is shown to exist. Previously a similar
 correspondence between the stress tensors was known and well established in
 the 70's. Using 
 recent results obtained in semiclassical stochastic gravity regarding exact
 definition and significance given to quantum fluctuations of the stress tensor,
we obtain this correspondence, which is fundamental to statistical
analysis of systems in curved spacetime. This is of immense importance, 
 in that the fluctuations of stress tensor play a similar central role 
 in  stochastic gravity, as  that of the stress tensor in classical and
 semiclassical gravity. A relation
 between the semiclassical and classical fluctuations therefore, gives insight
 to the mesoscopic phenomena for  gravitating systems and would further
enhance the perturbative analysis for cosmological spacetime and astrophysical 
objects, which is an expansive area of research.
Interestingly we see that the quantum fluctuations have a correspondence with
covaiances of pressure and density of the gravitating system in the stochastic
analysis.
\end{abstract}
\maketitle
In this letter, inspired by the correspondence between classical and quantum
 stress tensors which was established in 70's and is well known in literature
 \cite{ dodleson,zeldo2}, we attempt to establish a similar correspondence
 between the fluctuations of the two. This is
 an enhanced result that we obtain by using the  exact form of 
 noise kernel \cite{phillipnk} defining fluctuations of a quantum field,
 recently obtained in semiclassical stochastic gravity \cite{bei1},
 and relate it those of a perfect fluid stress tensor in the classical limit. 
Thus we show an equivalence of the two noise kernels in classicalized limit, 
 which opens up possibliltiy  for simplifying
mathematically more intricate quantum stress tensor issues and replacing them
 by simpler fluctuations of classical variables which are equivalent for
 statistical analysis. This is the first attempt to show any such
 relation between the quantum and classical fluctuations thus providing a
 link  between semiclassical and classical gravity to be treated equivalently
in the stochastic domain.  Of interest towards possible applications of 
such a correspondence would be the cosmological and astrophysical spacetimes   
  and  their analysis in both the equilibrium and non-equilibrium limits. 

Stress tensors describe the matter part of the Einstein's equations in
classical and semiclassical gravity.  
The issues of renormalization and regularization are important to get
 physically meaningful results for the quantum stress tensors \cite{birrel}.
 However the
 effect of such a renormalized quantum stress tensor on the 
 spacetime geometry is realized only through its average or expectation
 value which acts as a classical entity. 

On the other hand  classical matter described by scalar fields and perfect
 fluids etc. are commonly used to model matter content of the astrophysical
 objects or the universe \cite{dodleson,friedman, madsen}. 
 The correspondences between  stress tensors of scalar fields, perfect fluid 
 and expectation of  quantum stress tensor is a well established for
 relevant cases. Specially for a cosmological spacetime, the
relation between  expectation of quantum stress tensor with conformally
invariant scalar fields and a perfect fluid model is well known  and widely
 used \cite{ dodleson,zeldo2, lukash} . 

While the importance of the stress tensors, whether quantum or classical,
 for General Relativity cannot be
underestimated, fluctuations of the stress tensor are finding increasing 
applications in astrophysics and cosmology \cite{bei1, sukbh,sukratu,seema1}.  
 In this letter we are concerned about statistical fluctuations on the
 background spacetime and do not touch upon perturbations of the matter
 fields, as in a perturbative analysis. 
Such fluctuations of the quantum stress tensor have been defined
and given explicit form and meaning in semiclassical stochastic gravity 
\cite{phillipnk}.  
On the other hand, recently proposed classical stochastic gravity  shows 
 fluctuations of the classical stress tensor useful for analysing statistical
 properties of astrophysical objects \cite{seema2}. 
 We further attempt to obtain a correspondence between the two . 

The procedure that we follow to show this correspondence, takes the quantum
 fluctuations to classicalized form via transition from quantum scalar
fields to classical scalar fields (the reverse of quantizing the classical
 fields), then using the point splitting method as described later, and finally
using the relation between classical scalar fields and  pressure and energy
 density variables as in the perfect fluid desciption of matter.  
The key relation that enables us to obtain this very 
correspondence  is the  noise kernel for quantum 
fields.
 
A quantum stress tensor is obtained by quantizing
a scalar field ( here we consider a conformally invariant scalar field)
\be \label{eq:1}
T_{ab}(x)  =  \phi_{;a} \phi_{;b} - \frac{1}{2} g_{ab}
 \phi^{;c} \phi_{;c} + 1/6 (g_{ab} \Box - \nabla_a \nabla_b + G_{ab} )\phi^2
\ee
where $\phi$ is the classical scalar field which has to be quantized, to get a
 quantum stress tensor $\hat{T}_{ab}$.

Here we review the case studied in \cite{zeldo2} and work on Kasner metric,
 which decribes the anisoptropic 
 but homogenous cosmological spacetime, while our results also 
 extend to the FRW metric for the isotropic and homogenoeus case.
Thus we use the general form
\be \label{eq:metric}
ds^2 = - dt^2 + a^2(t) dx_1^2 + b^2(t) dx_2^2 + c^2(t) dx_3^2 
\ee 
where $a, b$ and $c$ are certain non negative functions of time ( including
the case where $a=b=c$, for FRW ). We are interested in the epoch where, in the
early universe, $ t_0 >> t_{Pl} $, such that  $t < t_0$, is the  time when
 there is 
 no particle production  thus for $ t = t_0$ one can
define a vacuum state correctly and particle production is switched on. 
Then for $t > t_0 $ one considers the expectation value of the quantum
stress tensor to have non-zero values. The epoch when the influence of the
 quantum particles on the metric becomes important, one can neglect quantum
effects and consider only the classicalized effects given by the expectation
of stress tensor, thus falling in the regime of 
\be
 G_{ab} = < \hat{T}_{ab} >
\ee
the semiclassical gravity.

We are interested thus in the case, where $t > t_0$ hence the classicalized 
system, where the formulation of quantum problem reduces to considering 
classical wave equation for the fields $\phi$. Then, only the following average
 values of the  quantum energy momentum tensor are different from
 zero and are given by
$<0| \hat{T}^0_0|0 > = - \epsilon , <0|\hat{T}^1_1|0> = <0|\hat{T}^2_2|0> =
 <0|\hat{T}^3_3|0> = p $. 
  The vacuum expectation
values given above diverge due to zero-point oscillations of the vacuum and are
 hence renormalized, giving renormalized values of the energy density
 and pressure \cite{zeldo2}. From now on we would address only the renormalized 
energy density and pressure. Thus one can imagine a  perfect fluid
given by $T^a_b =\mbox{ diag}\{-\epsilon, p, p, p \} $ to represent the matter
 part of the Einstein equations for this model.

 We begin with the two point noise kernel as has been obtained
in \cite{phillipnk} and attempt to show that it can be directly related to 
that of the perfect fluid case for a  Kasner and an FRW metric.

The semiclassical two point noise kernel is defined by
\begin{widetext}
\bea \label{eq:noise1}
8 \hat{N}_{abc'd'}(x,y) & = & < \{ \hat{T}_{ab}(x) - < \hat{T}_{ab}(x)> ,
 \hat{T}_{c'd'}(y) - <\hat{T}_{c'd'}(y) > \} > \nonumber \\
& & < \{\hat{T}_{ab}(x), \hat{T}_{c'd'}(y) \} > - 2 <\hat{T}_{ab}(x)>
<\hat{T}_{c'd'}(y) > 
\eea 
\end{widetext}
In \cite{phillipnk} point splitting  method in the above equation was used 
 to obtain the following expression for the noise kernel. 
\be \label{eq:noise2}
\hat{N}_{abc'd'} = Re\{ \bar{K}_{abc'd'} + g_{ab} \bar{K}_{c'd'} + g_{c'd'} 
\bar{K}'_{ab} + g_{ab} g_{c'd'} \bar{K} \}
\ee
where,
\bea \label{eq:nk}
& & 9 \bar{K}_{abc'd'}  =  4 ( \G_{;c'b} \G_{;d'a} + \G_{;c'a} \G_{d'b}) + 
\G_{;c'd'} \G_{;ab}  \nonumber \\
& & + \G \G_{;abc'd'}-2( \G_{;b} \G_{;c'ad'} + \G_{;a} \G_{;c'bd'} + \G_{;d'}
 \nonumber \\
& &  \G_{;abc'} + \G_{;c'} \G_{;abd'} ) + 2 ( \G_{;a} \G_{;b} R_{c'd'} +
 \G_{;c'} \G_{;d'} R_{ab} ) \nonumber \\
& & - ( \G_{;ab} R_{c'd'} + \G_{c'd'} R_{ab} ) \G + \frac{1}{2} R_{c'd'} R_{ab} \G^2 \nonumber \\
& & 36 \bar{K}'_{ab} = 8 (- \G_{;p'b} \tensor{\G}{_;_{p'}_a^{p'}}+ \G_{;a} 
\tensor{\G}{_;_{p'}_b^{p'} }) + 4 ( \G_;^{p'} \G_{;abp'} \nonumber \\
 & &  - \G_{;p'}^{p'}\G_{;ab} - \G_{abp'}^{p'} )-2 R'( 2 \G_{;a} \G_{;b} -
 \G \G_{;ab}) - \nonumber \\
& &  2 ( \G_{;p'} \G_;^{p'} - 2 \G \G_{;p'}^{p'} ) R_{ab} - R' R_{ab} \G^2
 \nonumber \\
& & 36 \bar{K} = 2 \G_{;p'q} \G_;^{p'q} + 4 ( \G_{;p'}^{p'} \G_{;q}^q +
 \G \tensor{\G}{_;^p_{q'}^{q'}} ) - \nonumber \\
& & 4 ( \G_{;p} \tensor{\G}{_;_{q'}^p^{q'}} 
 + \G_;^{p'} \tensor{\G}{_;_q^q_{p'}} ) R \G_{;p'} \G_;^{p'} + R' \G_{;p}
 \G^{;p} \nonumber \\
& &  - 2 (R \G_{;p'}^{p'} + R' \G_{;p}^{p}) \G + \frac{1}{2} R R' \G^2. 
\eea
where the Wightmann functions  denoted by $\G$ are defined as 
\be \label{eq:w}
\G(x,y) = <\{\hat{\phi}(x),\hat{\phi}(y) \}>.
\ee
 Point splitting is usually used for dealing with ill defined  quantum 
operators, which are in the form of $\hat{\phi}^2 $ \cite{birrel, bei1,
phillipnk}. The
 prescription is
 to take $\hat{\phi}(x) \rightarrow \hat{\phi}(x) \hat{\phi}(x') $, and use this
to evaluate the final expressions, which at the end are  consistently put back
in a much desired form by taking back $x' \rightarrow x $ for the functions 
and operators. 
However the quantum nature  of the operators here has  nothing
 to do with the mathematical artefact of using point splitting. Thus, we
 employ the same for classical case as well and get consistent 
results. This is discussed further. 
 
The Wightmann function given above by (\ref{eq:w}), in the classical 
limit  reduces to
\[ \G(x,y) \rightarrow G(x,y) =  <\phi(x) \phi(y)>,\] 
with $\phi$ classical. Replacing $\G$ by $G$ in equation (\ref{eq:nk}), gives 
us corresponding  noise kernel in the classicalized system.
The expression of such a noise kernel (classicalized)  thus obtained can be
 seen to match, with 
\be \label{eq:cT}
<T_{ab}(x) T_{cd}(y)> - <T_{ab}(x)><T_{cd}(y) > 
\ee
It can be (conversely) formally shown, that if one employs the point splitting
 procedure also for the classical fields given by the stress tensor 
(\ref{eq:1}) in the above expression 
we obtain a similar expression as that in (\ref{eq:noise2}) and
(\ref{eq:nk}), with $\G$ replaced by $G$. 
Hence one obtains the correspondence: $\hat{N}_{abcd}(x,y) \rightarrow
 N_{abcd}(x,y) $. We now work on the latter form.
  It can be directly verified from (\ref{eq:noise1}) that in the classical
limit,
\be \label{eq:fluid}
4\tensor{N}{^a_b^c_d}(x,y) =  < T^a_b(x) T^{c'}_{d'}(x)> - <T^a_b(x)><T^a_b(y)> 
\ee
Here the averages are over the classical field distribution. 

For the metric given by (\ref{eq:metric}) , 
$a(t), b(t) , c(t) $ the scalar field $\phi(t) $ and the fluid variables
$\epsilon(t), p(t) $ are homogeneous.  For such a case, only the
diagonal elements  of the averaged quantum stress tensor are important and
 give non-zero contribution as mentioned earlier. 
Further, using the relation between scalar fields and pressure/density,
one can easily see that the two corresponding forms of stress 
tensors are interchangeable and thus one can get the noise kernel in 
terms of perfect fluid variables here.
 This can be modeled by a stress tensor with diagonal non-zero elements 
 $T^0_0 = - \epsilon(t), T^1_1 = T^2_2= T^3_3= p(t) $ of the classical stress
 tensor  for homogeneous case.
The procedure we have followed here, takes 
$\hat{T}_{ab}$ (quantum)$ \rightarrow T_{ab}$ (classical $\phi$ ) $\rightarrow 
 T_{ab} $ (perfect fluid).
 
The perfect fluid, has collisons and microscopic effects
 which give rise to non-zero pressures. This enables one to consider the
 stress tensor as
 a random variable \cite{seema2,friedman}, and introduce a statistical
 treatment for the same, such that averages of
 fluid variables are well defined in the distributional sense.

 Putting this in (\ref{eq:fluid}) we get the following
 non-zero components of the noise kernel. 
\bea
4\tensor{N}{^0_0^0_0}(t,t') &=& Cov[ \epsilon(t),\epsilon(t')] \nonumber \\
4\tensor{N}{^0_0^i_i}(t,t') & = & Cov[\epsilon(t),p(t')] \nonumber \\
4\tensor{N}{^i_i^0_0}(t,t') & = & Cov[ p(t), \epsilon(t')] \nonumber \\
4\tensor{N}{^i_i^j_j}(t,t') & = & Cov[p(t),p(t')]
\eea  
Thus the covarainces of pressure and density capture the microscopic
nature of the fluid in a statistical description which are thus related
to quantum fluctuation inherently. 

It is known that the noise kernel is of central importance to stochastic
gravity, which is a step up from semiclassical gravity in the 
perturbative approach.  In addition to being used  in the perturbative
theory as noise, these fluctuations characterize  the  microscopic
effects in  homogeneous cosmological spacetime itself.
Thus  our results indicate that, fluctuations of quantum fields induce
 mesoscopic classicalized effects in the fluid 
description of the same model, given by covariances of pressures and energy
density in the background spacetime.

Here we have extended the correspondence between quantum and classical stress 
tensors to the fluctuations (two point)  of the same.
 We emphasise, that  this  holds  in the classical limit, where
 the  purely quantum effects
 can be ignored, and thus only the classicalized effect of quantum stress
 tensor is important. This is by no means a trivial case, since
 for semiclassical gravity, one always uses the expectation of the 
quantum stress tensor in the Einstein's equation.
 
The  noise kernel $\hat{N}_{abcd}$  due to quantum fluctuations, is 
 classical and stochastic in nature. One can thus raise a question about
the correspondence worked out here. The importance 
of what we have shown here, lies in realizing that the quantum sourced
noise due to quantum fluctuations can also be captured partially
  via the classical noise driven by perfect fluid model of matter for the
  homogenoeus Kasner and FRW metric. Hence a basic link between quantum and
 classical describtion of the fluctuations, on the lines of corresponding 
 stress tensors, has been established in the  sense of treating matter
fields and their properties in the two domains.
\begin{acknowledgments}
S.Satin is thankful to Bei Lok Hu for useful discussions. This work
was carried out as part of the project funded by Department of Science
and Technology (DST), India through grant no. DST/WoS-A/2016/PM/100
\end{acknowledgments}

\end{document}